\documentclass[onecolumn,showpacs]{revtex4}

\usepackage{graphicx}
\usepackage{dcolumn}
\usepackage{bm}

\input epsf

\begin{document}

\title{\Large Varying Speed of Light, Modified Chaplygin Gas and Accelerating Universe}

\author{\bf Anup Kumar Singha$^1$ and
Ujjal Debnath$^2$\footnote{ujjaldebnath@yahoo.com ,
ujjal@iucaa.ernet.in}}

\affiliation{$^1$Department of Mathematics, Calcutta Institute of Technology, Uluberia-711 316, India.\\
$^2$Department of Mathematics, Bengal Engineering and Science
University, Shibpur, Howrah-711 103, India.}

\date{\today}

\begin{abstract}
In this paper, we have considered a model of modified Chaplygin
gas in VSL theory with variable gravitational constant $G$. We
have shown that the evolution of the universe starts from
radiation era to phantom model. The whole evolution of the
universe has been shown diagramatically by using statefinder
parameters.
\end{abstract}


\maketitle

\section{\normalsize\bf{Introduction}}
Recently, the  matter  field  can  give  rise  to  an
accelerated  expansion [1]  for  the  universe stems  from  the
observational  data  regarding  the  luminosity-redshift
relation  of  type Ia  supernovae [2,3]  up  to  about $z\sim 1$.
This  matter  field  is  called  {\it quintessence matter}
(shortly Q-matter). This  Q-matter  can  behave  like a
cosmological constant [4,5]  by  combining  positive  energy
density  and negative  pressure. So there  must  be  this
Q-matter  either neglected  or  unknown  responsible  for  this
accelerated universe. At  the  present  epoch, a  lot  of works
has been done  to  solve  this  quintessence  problem and most
popular candidates  for  Q-matter  has  so  far  been a scalar
field having  a  potential  which  generates  a sufficient
negative pressure. The Q-matter behaves as a perfect fluid with a
barotropic equation of state and so some effort has been
investigated in determining its adiabatic index at the present
epoch [6, 7]. Chimento  et al [8] have showen that  a  combination
of dissipative  effects  such as  a bulk viscous  stress and  a
quintessence  scalar  field gives an accelerated expansion for
an  open  universe  ($k = -1$). Banerjee et al [9]  also  have
shown that it is possible to  have an accelerated  universe  with
BD-theory in Friedmann  model without  any  matter.\\

Another alternative candidate for Q-matter is exotic type of
fluid $-$ the so-called Chaplygin gas which obeys the equation of
state $p=-B/\rho, (B>0)$ [10], where $p$ and $\rho$ are
respectively the pressure and energy density. Subsequently, the
above equation was generalized to the form $p=-B/\rho^{\alpha},
0\le \alpha \le 1$ [11, 12] and recently it was modified to the
form $p=A\rho-B/\rho^{\alpha}, (A>0)$ [13, 14], which is known as
{\it Modified Chaplygin Gas}. This model represents the evolution
of the universe starting from the radiation era to the $\Lambda
CDM$ model.\\

The  possibility  that  the  speed  of  light  $c$  might  vary
has  recently  attracted  considerable  attention. In a
cosmological  setting, the  variations  in  $c$  have been shown
to  solve  the  cosmological  puzzles - the  horizon, flatness
and  Lambda  problems  of  big-bang cosmology. The variations of
velocity  of  light  can  also  solve  the quasi-lambda problems.
For  power-law  variations  in  the velocity  of light  with the
cosmological  scale  factors, Barrow  et al [15]  have shown
that   flatness   problem can be  solved. The  Machian VSL
scenario  in   which $c=c_{0}a^{n}$, introduced  by  Barrow [16]
has  significant advantages  to the   phase   transition
scenario  in   which the  speed  of light  changes  suddenly
from  $c^{-}$  to $c^{+}$, preferred by  Albrecht and  Magueijo
[17]. For changing $c$, the geometry of  the universe  is  not
affected. In classical electromagnetism the speed of light is
only constant in vacuum and it `varies' in dielectric media. The
changing $c$ breaks Lorentz invariance and conservation of
energy. The geometry of Universe is not affected by a changing
$c$. One can allowed a changing $c$ to do the job normally done
by {\it superluminal expansion}. For changing $c$, the
gravitational laws should be modified. The basic assumption is
that a variable $c$ does not induce corrections to curvature in
the cosmological frame and therefore, Einstein's equations,
relating curvature to stress energy are still valid. The reason
behind this postulate is that $c$ changes in the Local Lorentzian
frames associated with cosmological expansion. The effect is a
special relativistic effect and not a gravitational effect.
Therefore curvature should not be related with the variation of $c$.\\

The organization of the paper is as follows: In  section II, we
have studied field equations and its solutions due to VSL theory
for modified Chaplygin gas with varying $G$. In section III, we
have discussed whole stage of the evolution of the universe using
statefinder parameters. In the last section (i.e., section
IV) we give some remarks of this paper.\\

\section{\normalsize\bf{Basic Equations and Solutions}}

In VSL theory and varying $G$, the Einstein equations for FRW
model is

\begin{equation}
\frac{\dot{a}^{2}}{a^{2}}+\frac{k c^{2}(t)}{a^{2}}=\frac{8\pi
G(t)\rho }{3}
\end{equation}
and
\begin{equation}
\frac{\ddot{a}}{a}=-\frac{4\pi
G(t)}{3}\left(\rho+\frac{3p}{c^{2}(t)}\right)
\end{equation}

where $p$ and $\rho$ are the pressure and energy density
respectively. Here we consider modified Chaplygin gas in VSL
theory as

\begin{equation}
p=A\rho c^{2}(t)-\frac{B}{\left\{\rho
c^{2}(t)\right\}^{\alpha}}~~,~A>0,~~B>0,~~0\le\alpha\le1.
\end{equation}

The energy conservation equation incorporating time variation in
$c(t)$ and $G(t)$ is

\begin{equation}
\dot{\rho}+3\frac{\dot{a}}{a}\left(\rho+\frac{p}{c^{2}}\right)=-\frac{\dot{G}}{G}~\rho+\frac{3kc\dot{c}}{4\pi
Ga^{2}}
\end{equation}

Now assume, the velocity of light and $G$ are in power-law form
of scale factor as

\begin{equation}
c(t)=c_{0}a^{n}
\end{equation}
and
\begin{equation}
G=G_{0}a^{m}
\end{equation}

where $c_{0}$ and $G_{0}$ are both positive constants. Since we know that speed of light
decreases with time and $G$ increases with time, so that $n$ must be negative and $m$ must be positive.\\

Using equations (3), (4), (5) and (6) we have the solution of
$\rho$ as (for $k=0$)

\begin{equation}
\rho=\left[M~a^{-2n(1+\alpha)}+N~a^{-(m+3A+3)(1+\alpha)}
\right]^{\frac{1}{1+\alpha}}
\end{equation}

where $M=\frac{3B}{(3+3A-2n+m)c_{0}^{2(1+\alpha)}}$,
$N=\frac{c_{1}}{g_{0}^{1+\alpha}}$ and $c_{1}$ is a constant.\\

Putting the value of $\rho$ from (7) in (2), we have

\begin{equation}
\int
\frac{a^{\frac{3A+1}{2}}~da}{\left[M~a^{(3-2n+m+3A)(1+\alpha)}+N
\right]^{\frac{1}{2(1+\alpha)}}} =c_{2}~t
\end{equation}

where $c_{2}=\sqrt{\frac{8\pi G_{0}}{3}}$.\\

The solution of scale factor $a(t)$ has the form

\begin{equation}
a^{\frac{3(1+A)}{2}}~_{2}F_{1}[\frac{1}{2(1+\alpha)},\frac{3(1+A)}{2(3+3A+m-2n)},1+\frac{3(1+A)}{2(3+3A+m-2n)},
-\frac{M}{N}~a^{3+3A+m-2n} ]
=\frac{3}{2}(1+A)c_{2}N^{\frac{1}{2(1+\alpha)}}~t
\end{equation}

where $_{2}F_{1}$ is the hypergeometric function.\\

Now for small value of scale factor $a(t)$ i.e., for early
universe, $\rho$ is very large and corresponds to the universe
dominated by an equation of state $p=A\rho c^{2}$.\\

Also for large value of scale factor $a(t)$ i.e., for late
universe, the relation between $p$ and $\rho$ satisfies the
equation of state
\begin{equation}
p=-\left(1+\frac{m-2n}{3} \right)\rho c^{2}
\end{equation}

So evolution of the universe starts from radiation era ($A=1/3$)
to phantom model. From observations, we can conclude that $m\le 2(n+1)$.
For $\Lambda CDM$ model we have $m=2n$.\\

\section{\normalsize\bf{Statefinder Parameters and Evolution of the Universe}}

In 2003, V. Sahni et al [18] proposed a pair of parameters
$\{r,s\}$, called {\it statefinder} parameters. In fact,
trajectories in the $\{r,s\}$ plane corresponding to different
cosmological models demonstrate qualitatively different behaviour.
The above statefinder diagnostic pair has the following form:

\begin{equation}
r=\frac{\dddot{a}}{aH^{3}}~~~~\text{and}~~~~s=\frac{r-1}{3\left(q-\frac{1}{2}\right)}
\end{equation}

where $H\left(=\frac{\ddot{a}}{a}\right)$ and
$q~\left(=-\frac{a\ddot{a}}{\dot{a}^{2}}\right)$ are the Hubble
parameter and the deceleration parameter respectively. These
parameters allow us to characterize the properties of dark energy.
Trajectories in the $\{r,s\}$ plane corresponding to different
cosmological models, for example $\Lambda$CDM model
diagrams correspond to the fixed point $s=0,~r=1$.\\

The expressions for $q$, $r$ and $s$ are in the following:

\begin{equation}
q=\frac{1+3A}{2}-\frac{3B}{2(\rho c^{2})^{1+\alpha}}
\end{equation}

\begin{eqnarray*}
r=-\frac{1+3A}{2}+\frac{3B}{2(\rho
c^{2})^{1+\alpha}}-\frac{3}{2}~m~G_{0}a^{m}\left(1+Ac^{2}-\frac{B}{(\rho
c^{2})^{1+\alpha}} \right)~~~~~~~~~~~
\end{eqnarray*}
\begin{equation}
-\frac{a}{2}~\left(1+3A+\frac{3\alpha B}{(\rho c^{2})^{1+\alpha}}
\right)\frac{d\rho}{da}+3n\left(A\rho-\frac{B}{\rho^{\alpha}c^{2\alpha+2}}
\right)
\end{equation}
and
\begin{eqnarray*}
s=\frac{1}{3\left(A-\frac{B}{(\rho c^{2})^{\alpha}}
\right)}\left[-(1+A)+ \frac{B}{(\rho
c^{2})^{1+\alpha}}-m~G_{0}a^{m}\left(1+Ac^{2}-\frac{B}{(\rho
c^{2})^{1+\alpha}} \right)\right.
\end{eqnarray*}
\begin{equation}
\left. -\frac{a}{3}~\left(1+3A+\frac{3\alpha B}{(\rho
c^{2})^{1+\alpha}}
\right)\frac{d\rho}{da}+2n\left(A\rho-\frac{B}{\rho^{\alpha}c^{2\alpha+2}}
\right)  \right]
\end{equation}

The expressions of $r$ and $s$ are very complicated, so we can
not found the relation between $r$ and $s$ in closed form.\\

Fig. 1 shows the variation of $q$ against $a$ for $A=1/3,
~\alpha=1/2,~B=1,~n=-1/2,~m=1$. The plot of $q$ against $a$
reveals that the deceleration parameter indeed has a sign flip in
the desired direction and indicates an early deceleration ($q>0$)
followed by a late time acceleration ($q<0$) of the universe. From
fig. 2 we see that the curve starts from the radiation era and
goes asymptotically to he dust model for $s>0$ and $r>7$. But for
$r<7$, the portion of the curve represents the evolution from dust
state ($s=-\infty$) to phantom model ($s>0$) and this curve
crosses $\Lambda$CDM stage ($r=1, s=0$). Thus the total curve
represents the evolution of the universe starting from radiation
era to phantom model.\\

\begin{figure}
\includegraphics[height=2.7in]{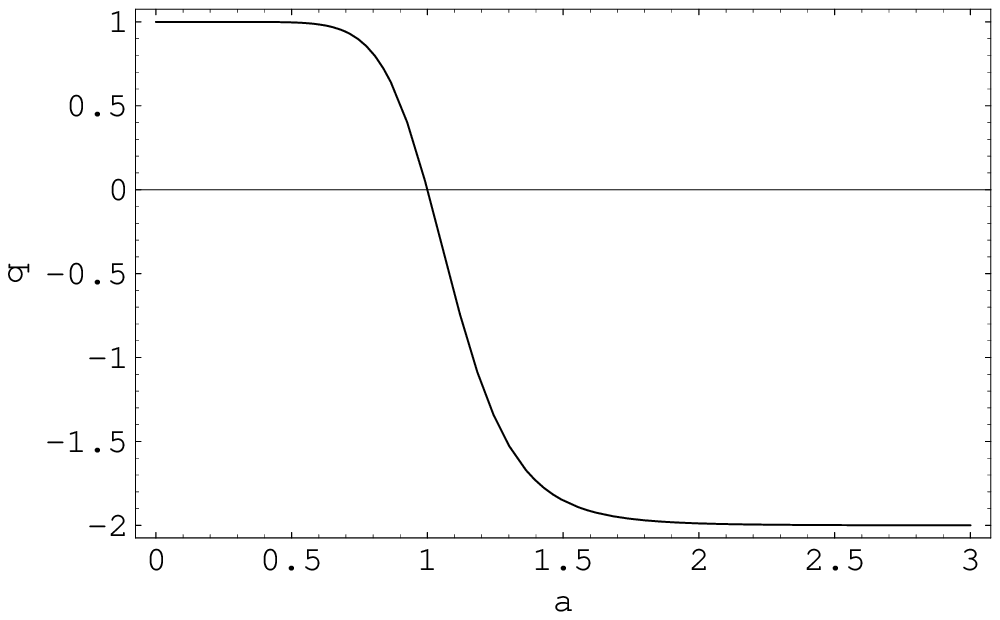}\\
\vspace{1mm} Fig.1\\

\vspace{7mm}
\includegraphics[height=2.7in]{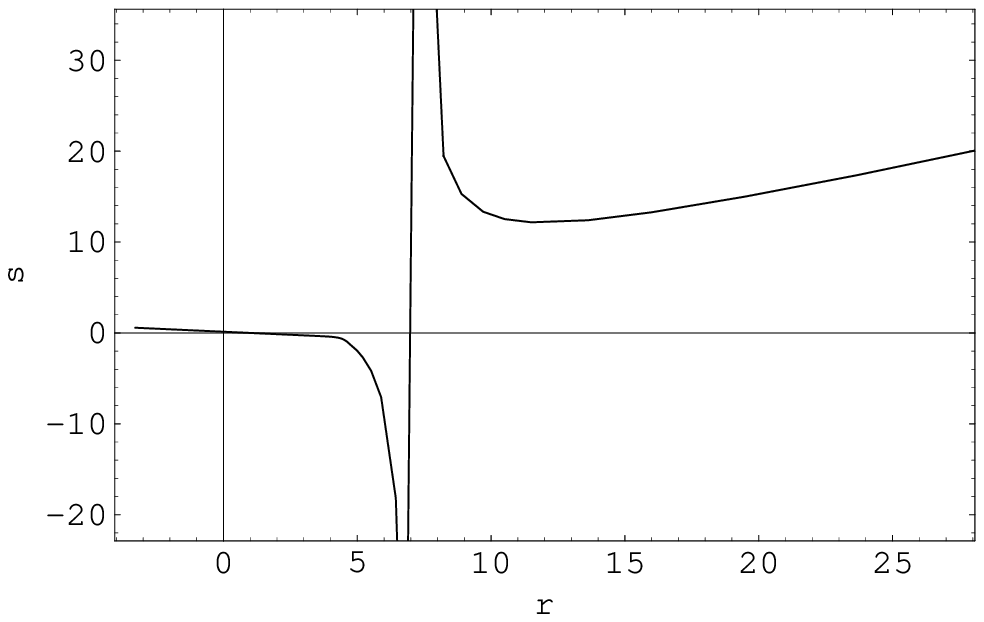}\\
\vspace{1mm} Fig.2\\

\vspace{7mm} Fig.1 shows variation of $q$ against $a$ and Fig.2
shows variation of $s$ against $r$ for
$\alpha=1/2,~A=1/3,~B=1,~n=-1/2,~m=1$.\vspace{6mm}

\end{figure}

\section{\normalsize\bf{Discussions}}

In this work, we have proposed a model for modified Chaplygin gas
in VSL theory with variable $G$. In this model, we have found the
evolution of the universe starts from radiation era to phantom
model. If $m=n=0$ i.e., $c=$constant and $G=$constant then we can
recover our previous results [14]. Thus variable $c$ and $G$ can
drive the evolution of the universe from radiation to phantom
stages instead of $\Lambda$CDM stage. From equation (10) we see
that for $\Lambda$CDM model $m=2n$ and for phantom model $m\le
2(n+1)$. If $G=$constant or $c=$constant then any one of variable
$G$ and $c$ can drive the evolution of the universe into the
phantom model. Thus variable $G$ or $c$ has significant role for
the evolution of the universe in large stages.\\

{\bf References:}\\
\\
$[1]$  N. A. Bachall, J. P. Ostriker, S. Perlmutter, P.
J. Steinhardt, {\it Science} {\bf 284} (1999) 1481.\\
$[2]$  S. Perlmutter et al 1998 {\it Nature} (London) {\bf 391}
51; 1999 {\it Astrophys  J} {\bf 517} 565.\\
$[3]$  A. G. Riess   et al 1998 {\it Astrophys  J} {\bf 116} 1009;
P. M. Garnavich   et al  1998 {\it Astrophys  J} {\bf 509} 74.\\
$[4]$  R. R. Caldwell, R. Dave  and  P. J. Steinhardt, {\it Phys.
Rev. Lett.} {\bf 80}, 1582 (1998) [{\it astro-ph}/9708069].\\
$[5]$  V. Sahni, A. A. Strarobinsky, {\it Int. J. Mod. Phys. D}
{\bf 9} 373 (2000).\\
$[6]$  L. Wang, R. Caldwell, J. P. Ostriker  and  P. J.
Steinhardt, {\it Astrophys. J.} {\bf 530} 17 (2000).\\
$[7]$  I. Waga and P. M. R. Miceli, {\it Phys. Rev. D} {\bf 59}
103507 (1999).\\
$[8]$  L.  P. Chimento, A.  S. Jakubi    and D. Pavo'n, {\it Phys.
Rev. D} {\bf 62} 063508 (2000) (Preprint  {\it astro-ph}/005070).\\
$[9]$ N. Banerjee  and  D. Pavo'n, {\it Phys. Rev. D}
{\bf 63} 043504 (2000).\\
$[10]$ A. Kamenshchik, U. Moschella and V. Pasquier, {\it Phys.
Lett. B} {\bf 511} 265 (2001).\\
$[11]$ V. Gorini, A. Kamenshchik and U. Moschella, {\it Phys. Rev.
D} {\bf 67} 063509 (2003); U. Alam, V. Sahni , T. D. Saini and
A.A. Starobinsky, {\it Mon. Not. Roy. Astron. Soc.} {\bf 344}, 1057 (2003).\\
$[12]$ M. C. Bento, O. Bertolami and A. A. Sen, {\it Phys. Rev. D}
{66} 043507 (2002).\\
$[13]$ H. B. Benaoum, {\it hep-th}/0205140.\\
$[14]$ U. Debnath, A. Banerjee and S. Chakraborty, {\it Class.
Quantum Grav.} {\bf 21} 5609 (2004).\\
$[15]$  J. D. Barrow  and  J. Magueijo, {\it Phys. Lett. B} {\bf
443} 104 (1998); {\it Phys. Lett. B} {bf 447} 246 (1999); {\it
Class. Quantum  Grav.} {\bf 16} 1435  (1999). \\
$[16]$  J. D. Barrow, {\it Phys. Rev. D} {\bf 59} 043515 (1999).\\
$[17]$  A. Albrecht  and  J. Magueijo, {\it Phys. Rev. D} {\bf 59}
043516 (1999). \\
$[18]$ V. Sahni, T. D. Saini, A. A. Strarobinsky and U. Alam, {\it
JETP Lett.} {\bf 77} 201 (2003).\\

\end{document}